\hoffset = -5mm
\magnification=1150
\vglue1cm \line {\vbox{\halign{\hfill#\hfill\cr Nice INLN 00/19 \cr}}
\hfill\vbox{\halign{\hfill#\hfill\cr July 2000\cr}}} \vglue3cm

\centerline { {\bf ON A FATE}}\smallskip
\centerline { {\bf OF}}\smallskip
\centerline { {\bf HOT QUANTUM FIELD THEORIES }}
\smallskip  \smallskip  \bigskip
\bigskip\bigskip\bigskip\bigskip \centerline {{${\rm{Bernard\
Candelpergher^\star\ and\  Thierry\ Grandou^\dagger}} $ }}
\bigskip\medskip \centerline {$\star$ {\it Laboratoire J.A. Dieudonn\'e,
UMR CNRS 6621; UNSA Parc Valrose, 06108 Nice, France }} \smallskip
\centerline { $\dagger${\it Institut Non Lin\'eaire de Nice UMR CNRS
6618; 1361, Route des Lucioles, 06560 Valbonne, France}}\centerline
{e-mail:grandou@inln.cnrs.fr}
\bigskip\bigskip\medskip\bigskip\bigskip \centerline{\bf ABSTRACT}
\bigskip\medskip
It is argued that for hot quantum fields, the necessary effective  
perturbation theories may be based on a
resummation procedure which, contrarily to the zero temperature  
case, differs substantially from the one
ordinarily in use. Important differences show up in the infrared  
sector of hot quantum field theories.
\bigskip\bigskip\noindent
PACS: 12.38.Cy, 11.10.Wx

\bigskip\bigskip\medskip\bigskip\bigskip
\vfill\eject
The series $\sum_{n=0}^\infty (-z)^n $, with $z$ a complex number
 defines an analytic function of $z$, and in a disc of radius one  
centered at the
origin $D(0,1)$ one has
$$\sum_{n=0}^\infty (-z)^n={1\over 1+z}\ ,\ \ \ \ z\in  
D(0,1)\eqno(1)$$ The function
$$z\longrightarrow f(z)={1\over 1+z}\ ,\ \ \ \ z\in C\!\!\!\!  
|-\{-1\}\eqno(2)$$ defines an analytic function of
$z$
in the whole complex plane but
for a simple pole at $z=-1$. It can be considered the analytical  
continuation of the series (1), and then, its
pole is
reminiscent, in view of a classical theorem, of the convergence  
radius of the above series. Indeed, it is well
known that in the past, relations between series and functions that  
express
their resummations have been the matter of intense and passionate  
controversies [1]. Euler, for example, used
the
above equations to derive such a counter-intuitive result as  
$$1-1+1-1+1-1+\ ..={1\over 2}$$
\medskip
In loop calculations of (renormalizable) Quantum Field Theories  
(QFT's), things are more involved
since the $z$'s of (1) and (2) are no longer complex $c$-numbers,  
but complex valued one dimensional
distributions
in some conveniently choosen variable, such as the energy $k_0$ or  
the virtuality $K^2=k_0^2-k^2$ of the looping
momenta. Let us be more specific and consider the case of QED. The  
bare photon Feynman propagator can be written
as
$$D^{(0)}_{\mu\nu}(K)={i\over  
K^2+i\varepsilon}\left(d^T_{\mu\nu}(K)+d^L_{\mu\nu}(K)\right)\eqno(3)$$where  
the
transverse (T) and
longitudinal (L)
"Klein-Gordon divisors", $d^{L,T}_{\mu\nu}(K)$ are dependent on the  
gauge condition choosen to quantize the
theory. The fully dressed (renormalized) photonic propagator,  
${}^\star D_{\mu\nu}(K)$ obeys a Schwinger-Dyson
equation
$${}^\star  
D_{\mu\nu}(K)=D^{(0)}_{\mu\nu}(K)+D^{(0)}_{\mu\sigma}(K)\Pi^{\sigma\lambda}(K){}^\star  
D_{\lambda\nu}(K)\eqno(4)$$where
$\Pi(K)$ is consistently interpreted as the full (renormalized)  
$1PI$ photonic polarization tensor. This equation
is satisfied
by the
series $$\sum_{n=0}^\infty  
D^{(0)}(K)\left(-iD^{(0)}(K)\Pi(K)\right)^{n}\eqno(5)$$where Lorentz  
indices have been
omitted for short. Now, in the course of practical loop  
calculations, whenever dressed propagators are to be
used,
one rather relies on the so called
resummed expression
$${}^\star  
D_{\mu\nu}(K)=\left((D^{(0)}(K))^{-1}-\Pi(K)\right)^{-1}_{\mu\nu}={i\over
K^2+i\varepsilon}\left(Z_3d^T_{\mu\nu}(K)+d^L_{\mu\nu}(K)\right)\eqno(6)$$Renormalization  
of $\Pi_{\mu\nu}(K)$ is
understood so as to make the supports of (5) and (6) coincide over  
some domain of $K$, the same way (1) and (2)
coincide over $D(0,1)$.
Equation (6) exhibits the features of photon wave function  
renormalization (the constant $Z_3$, related to the
function $\Pi(K)$), the absence of (genuine) photon mass  
renormalization, the dynamically decoupled
longitudinal component .. all standart, textbook material results.  
This means that, roughly speaking, the
analytical structures of (5) and (6) will be comparable; in  
particular, excluding any discussion of, otherwise
spurious Landau's ghosts [2], poles in the $k_0$-complex plane are  
the same, located at $k_0=\pm k
(1-i\varepsilon)$. In loop calculations, where the one dimensional  
distribution character of both (5) and (6)
naturally comes about, it is therefore reasonable to expect that the
series representation (5) and its resummed expression (6) will lead  
to the same results. Indeed this very
question
seems to have been totally absent of the existing literature, and  
whenever
necessary, it is the form (6) for the fully dressed propagator  
which has always been used. This may be
why it is only very recently that an explicit comparison of results  
obtained through forms (5) and (6) has
been achieved at zero temperature. And effectively, calculations  
performed in the case of two dimensional
QED,
have shown that the forms (5) and (6) can be used with equal results [3].
\bigskip

In this letter, we take advantage of a recent, rather involved  
analyzis [4], to emphasize that,
contrarily to what happens at $T=0$, the above alluded equivalence  
of (5) and (6) totally breaks down at finite,
high temperature $T$: The series representation for the dressed  
propagator and its formally resummed expression
are not, between them, in such a relation as the series (1) and the  
function (2), or equivalently, at $T=0$, the
full propagator (6) and its series representation (5). Plugged into  
loop calculations, they do not yield the
same
results in general.
This rahter unexpected fate, is specific to the thermal context,  
and could involve interesting issues, some of
them being sketched in our conclusion.
\bigskip

At high temperature $T$ and small coupling constant $e$, there are  
compelling reasons for using
effective propagators such as (6), whenever one is calculating  
physical processes involving soft energy scale
fluctuations,
of order $eT$, [5]. A famous example is provided by the damping  
rate calculation of a
fermion with mass $M$ and four-momentum $(E,{\vec{p}})$, travelling  
through a thermalized plasma. The relevant
object is the discontinuity in $k_0$ (hereafter denoted by disc) of  
the effective propagator
$${}^\star D_{\mu\nu}(k_0,k)={K^2\over k^2}{i{\cal{P}}^L_{\mu\nu}(K)\over
K^2-\Pi^{HTL}_L(k_0,k)}+{i{\cal{P}}^T_{\mu\nu}(K)\over
K^2-\Pi^{HTL}_T(k_0,k)}=\sum_{L,T}\  
i{\cal{P}}^{L,T}_{\mu\nu}(k_0,k){}^\star\Delta_{L,T}
(k_0,k)\eqno(7)$$where we have introduced the customary projectors  
onto longitudinal (physical at
$T\neq 0$!)
and transverse degrees of freedom, taken in Coulomb gauge, and  
where the superscript $HTL$ (hard thermal loop)
is to remind that only the leading, order $e^2T^2$, part of the  
functions $\Pi_{L,T}$ is retained; this
superscript hereafter omitted for short, is understood in the  
sequel. In the fermionic mass shell limit, the
fermionic damping rate can be written as $${}^\star\gamma (E,
p)={e^2T\over 2\pi }\int^{k_M}_{k_m} k{\rm d}k\int_{-k}^{+k}{{\rm  
d}k_0\over
2\pi k_0}
\left( {\rm{disc}}{}^\star\Delta_L
(k_0,k)+(1-{k_0^2\over k^2})\  
{\rm{disc}}{}^\star\Delta_T(k_0,k)\right)\eqno(8)$$where $k_M$ is  
some
upper cut off
irrelevant to our concern, whereas the lower limit is essentially  
given by the fermionic
mass shell condition $2k_m\simeq E-{\sqrt{p^2+M^2}}$. In (8), a  
factor of $T/k_0$ stands for the infrared (IR)
most part of the bosonic statistical factor expansion
$$1+n(k_0)={T\over k_0}+{1\over 2}+{\cal O}({k_0\over  
T})\eqno(9)$$Finally, one has also
$${\rm{disc}}{}^\star\Delta_{L,T}(k_0,k)=\lim_{\eta=0^+}{}^\star\Delta_{L,T}(k_0+i\eta,k)-{}^\star\Delta_{L,T}(k_0-i\eta,k)\eqno(10)$$
In the literature, this is often written as twice the imaginary  
part of the ${}^\star\Delta_{L,T}(k_0,k)$
distributions, and this is correct provided the determination of  
the logarithmic function entailed in the
functions $\Pi_{L,T}(k_0,k)$ is appropriately choosen. The  
discontinuity in $k_0$ should otherwise be
considered the fundamental object [6].\smallskip\noindent

 Cauchy's theorem allows a straightforward
evaluation of the
$k_0$-integrals, $i.e$, the so called energy
sum rules
$$\int_{-k}^{+k}{{\rm d}k_0\over 2\pi k_0}\
{\rm{disc}}{}^\star\Delta_L(k_0,k) = {m^2\over
k^2(k^2+m^2)}-{2Z_L(k)\over \omega_L(k)}
\eqno(11)$$where $m$ is the Debye screening mass, the value of  
which only, disriminating between the two QED and
QCD cases: Within standart notations,
$$m^2_{QED}={e^2T^2\over 6}\ ,\ \ \ \
\ m^2_{QCD}={g^2T^2\over 6}(N_c+{N_f\over 2})\eqno(12)$$
In {\it{Linear Response Approximation}}, $\omega_{L,T}(k)$
satisfy the dispersion relations $$\omega_{L,T}^2(k)=k^2+{\cal{R}}e\
\Pi_{L,T}\left(\omega_{L,T}(k),k\right)\eqno(13)$$and the constant  
$Z_{L,T}(k)$,
are residues at effective propagator's time-like poles  
$(\pm\omega_{L,T}(k),k)$ with
$\omega_{L,T}(k)$ enjoying gauge independent determinations [7]. By  
the fermionic mass shell, $k_m$
tends to zero and the low $k$ ($k/m<<1$) behaviours of the relevant  
energy sum rules become a crucial issue.
If there is no problem with (11)
which behaves like $4/5m^2$ in the IR low $k$ regime, this is not  
so with the transverse
contribution, in particular with the sum rule
$$\int_{-k}^{+k}{dk_0\over 2\pi k_0}\  
{\rm{disc}}{}^\star\Delta_T(k_0,k)={1\over
k^2}-{2Z_T(k)\over \omega_T(k)}\simeq {1\over k^2}-{3\over  m^2}\  
,\ \ \ k/m<<1\eqno(14)$$Put into
(8), this sum rule is at the origin of a famous, eleven years old  
problem arising both in thermal QED and
QCD,
with an IR singularity plaguing
the result for the fermionic damping rate, when taken at mass shell [8]
$${}^\star\gamma (E,
p)_{|_{E={\sqrt{p^2+M^2}}}}\simeq{e^2T\over 2\pi }\int^{k_M}_{k_m}  
k{\rm d}k\left({1\over k^2}+..\right)=\ {e^2
T
\over
2\pi}\ln({2k_M\over E-{\sqrt{p^2+M^2}}})+..\eqno(15)$$Physically, this
is interpreted as due to unscreened magnetostatic (transverse)  
modes. In the case of QED, it is
known that no magnetic mass can be invoked to cure this IR disease [9],
whereas more room could be left in the case of QCD. This latter  
possibility, indicated first by lattice
simulations, has received an independent new support from another  
recent approach
[10], but at a lower number of space-time dimensions. According to a
recent opinion, though, the magnetic mass of thermal QCD, on the  
order of $g^2T$, could be too small to act as
efficient enough an
IR cut off [11], and in any case, the issue for QED is left the  
same.\medskip
We therefore analyze the problem on the basis of the series  
representation for the effective propagator
$${}^\star\Delta_{L} (K)=\sum_{N=0}^\infty \Delta^{(N)}_{L}(K)=
{K^2\over k^2}\sum_{N=0}^\infty
\left({1\over K^2+i\varepsilon}\right)^{N+1}
\left(\Pi_{L}(K)\right)^N$$
$${}^\star\Delta_{T} (K)=\sum_{N=0}^\infty \Delta^{(N)}_{T}(K)=
\sum_{N=0}^\infty
\left({1\over K^2+i\varepsilon}\right)^{N+1}
\left(\Pi_{T}(K)\right)^N\eqno(16)$$
 The same fermionic damping rate, evaluated at
mass
shell, is now obtained as
$${e^2T\over 2\pi
}\int_{k_m}^{k_M} {k{\rm d}k}\ \sum_{N=0}^\infty\int_{-k}^{+k}{{\rm  
d}k_0\over 2\pi
k_0} \left( {\rm{disc}}\Delta_L^{(N)}(k_0,k) +(1-{k_0^2\over k^2})\
{\rm{disc}}\Delta_T^{(N)}(k_0,k)\right)
\eqno(17)$$ Though in no way compulsory, the usual relation  
$\sum_N\Delta_{L,T}^{(N)}={}^\star\Delta_{L,T}$ may be
assumed: An interesting translation of the problematics is thus  
obtained in terms of a comparison between the
possible two  sequences of a sum ($N$) and an integral ($k_0$)  
operations involved in the loop calculation, $i.e$,

$$\sum_{N=0}^\infty\ \int_{-k}^{+k}{{\rm d}k_0\over 2\pi
k_0} \left( {\rm{disc}}\Delta_L^{(N)}(k_0,k) +(1-{k_0^2\over k^2})\
{\rm{disc}}\Delta_T^{(N)}(k_0,k)\right)
\eqno(18)$$

$$\int_{-k}^{+k}{{\rm d}k_0\over 2\pi
k_0}\ \sum_{N=0}^\infty \left( {\rm{disc}} \Delta_L^{(N)}(k_0,k)  
+(1-{k_0^2\over k^2})\
{\rm{disc}}\Delta_T^{(N)}(k_0,k)\right) \eqno(19)$$This is why,  
while the latter sequence is explicitly
relevant of
the {\it{Resummation Program}} ({\it{RP}}) [5], the former has been  
dubbed a {\it{Perturbative Resummation}}
scheme ({\it{PR}})  of the leading thermal effects [4]. Now,  
concerning the {\it{PR}} sequence (18) of the sum
and
integral operations,
the following results can be demonstrated in full rigour [4]

 $$\sum_{N=0}^\infty \ \left(\int_{-k}^{+k}{{\rm d}k_0\over 2\pi
k_0} (1-{k_0^2\over k^2})\ {\rm{disc}}\Delta_T^{(N)}(k_0,k)\  
\equiv\  0\right) = 0\eqno(20)$$

$$\sum_{N=0}^\infty\ \left(\int_{-k}^{+k}{{\rm d}k_0\over 2\pi
k_0}\ {\rm{disc}}\Delta_L^{(N)}(k_0,k)\ =\ {1\over k^2}(-{m^2\over  
k^2})^N\right)={m^2\over
k^2(k^2+m^2)}\ , \
{m\over k}\in D(0,1)
\eqno(21)$$with of course, different results for the damping rate  
and related physical pictures. In
particular, the familiar physical
picture
of unscreened transverse modes totally disappears in view of (20),  
and one has
$$\gamma(E,
p)_{|_{E={\sqrt{p^2+M^2}}}}=\gamma_L(E,p)_{|_{E={\sqrt{p^2+M^2}}}}={e^2T\over  
2\pi
}\int_{k_m}^{k_M} {k{\rm d}k}\ \sum_{N=0}^\infty {1\over  
k^2}(-{m^2\over k^2})^N\eqno(22)$$ In the region
$|{\vec{k}}|\geq m$, the series can be resummed and one gets a  
simple, unambiguous IR-finite result
for the leading order
damping rate, $\gamma(E, p)=(e^2T/4\pi)\ln 2$, which at leading  
order $e^2T$, retains no dependence on the upper
bound of integration $k_M$. The related physical picture is the one  
of a moving fermion experiencing the
thermalized medium it crosses, by emitting and re-absorbing a  
photon which probes the medium over distances that
do not
exceed
the Debye screening length $r\sim m^{-1}$. Now, choosing to  
analytically continue the original
series down to arbitrarily small $k$-values, one can observe that  
the same IR logarithmic singularity
as found in (15), is obtained along the {\it{PR}} sequence also,  
but attached this time to the longitudinal rather
than transverse
degrees of freedom.\smallskip
Getting back to (18) and (19), that are the very matter of our  
comparison, we have respectively
$$\sum_{N=0}^\infty\ \int_{-k}^{+k}{{\rm d}k_0\over 2\pi
k_0} \  {\rm{disc}}\Delta_L^{(N)}(k_0,k) = {m^2\over k^2(k^2+m^2)}\  
,\ \ \ \ \ {m\over
k_M}\leq{m\over k}\leq 1
\eqno(23)$$

$$\int_{-k}^{+k}{{\rm d}k_0\over 2\pi
k_0}\ \sum_{N=0}^\infty \  {\rm{disc}} \Delta_L^{(N)}(k_0,k)  =   
{m^2\over k^2(k^2+m^2)}-2{Z_L(k)\over
\omega_L(k) }\ ,\
\ \ \ \ {m\over k_M}\leq{m\over k}\leq \infty\eqno(24)$$whereof we  
see that the sum/integral
lack
of
commutation is not just in the domain extension, from $ m/k\leq 1 $  
to $ m/k\leq
\infty
$, as it is at $T=0$ (between (5) and (6) or equivalently between  
(1) and (2)), but more strikingly, in the
appearance of extra terms, associated to time-like poles and  
residues. In fact, this new feature is entirely due
to analytical peculiarities of one loop thermal polarization  
functions $\Pi_{L,T}$, that are at sharp
variance with
their $T=0$ renormalized counterparts [4,12]. Up to possible  
analytical continuations, our main statement can
therefore be given the concise expression,
$$[\sum_{N=0}^\infty\ ,\ \int_{-k}^{+k}{{\rm d}k_0\over 2\pi
}\  ]\ {\rm{disc}} \Delta^{(N)}_{L,T}\ \varphi(k_0)=
\int_{-\infty}^{+\infty}{\rm
d}k_0 \left(\delta(k^2-{k^2\over K^2}
{\cal{R}}e\Pi_{L}),\delta (K^2-{\cal{R}}e\Pi_{T})\right)\  
\varphi(k_0)\eqno(25)$$with
$\Delta^{(N)}_{L,T}$ as given in (16), and where one has [5]
$$\delta (k^2-{k^2\over  
K^2}{\cal{R}}e\Pi_{L}),\delta\left(K^2-{\cal{R}}e\Pi_{T}\right) =
Z_{L,T}(k)\bigl\lbrace\delta(k_0-\omega_{L,T}(k))-\delta(k_0+\omega_{L,T}(k))\bigr\rbrace\eqno(26)$$
In (25), $\varphi (k_0,k)$ stands for any suitable test function,  
like $k_0/k^2$ in next equation (27), or
$1/k_0$ in previous equation (24) for example, and could as well  
include the full, unexpanded statistical
factor
$1+n(k_0)$ of (9). Note that the $k_0$- integration range is not  
the same in either sides of (25): This should
come as no surprise if we keep in mind that in virtue of (10), the  
discontinuities in $k_0$ of
${}^\star\Delta_{L,T}$
involve the boundary values, above and below the cut from $-k$ to  
$+k$, of functions that are analytic in the
whole $k_0$-
complex plane, but for poles that have to be picked up, away from  
the integration range $[-k,+k]$.
It is worth observing that, however astounding
a result (20) may appear at first sight,  it comes in full  
consistency with the sum/integral
broken
commutativity,
such as specified by (25): Relying
on (14) and the extra sum rule, $$\int_{-k}^{+k}{dk_0\over 2\pi k_0  
}\ ({k_0\over k})^2\
{\rm{disc}}{}^\star\Delta_T(k_0,k)={1\over k^2}\  
\left(1-2\omega_T(k)Z_T(k)\right)\eqno(27)$$
the
transverse part of (19) is found to be

$$\left({1\over k^2}-2{Z_T(k)\over
\omega_T }\right)-{1\over k^2} \left( 1-2\omega_TZ_T \right) =  
{1\over k^2}-{1\over k^2}+2({\omega_TZ_T\over
k^2}-{Z_T\over
\omega_T })\eqno(28)$$
Now, dropping contributions attached to the poles, in view of (25),
we see that the right hand side of (28) is effectively zero, as  
$1/k^2-1/k^2$, and this is nothing but (20).
Note also that both in the ultraviolet regime, where thermal  
effects fade away, and in the zero temperature limit,
it is straightforward to verify that (25) displays the usual $T=0$  
commutativity of the
sum/integral operations, because of the exponential suppression of  
$Z_{L,T}$ residues at high enough $k/m\sim
k/eT
$ ratios. Eventually, one can easily understand and check that  
relations of type (25) apply indeed to a large
variety of thermal
situations, and can be derived in models, simpler
than the presently considered cases of QED and QCD [12]: It appears  
as a genuine fate of hot quantum field
theories and models, whenever $HTL$, that is, leading thermal  
fluctuations are present.
  \bigskip

 At zero temperature,
there is neither practical nor theoretical interest in dealing with  
a full propagator series representation. At
non zero temperature, things are very different. First, whenever  
required by the thermal context, the effective
propagators series representations provide a consistent way of  
resumming leading thermal
effects of hot quantum field theories. Now, the {\it{Perturbative  
Resummations}} so achieved are found at clear
cut
variance with the Resummation Program results. The differences are  
entirely encoded in pole terms and
associated residues, that are non perturbative and infrared  
sensitive objects: In the general case, {\it{RP}} and
{\it{PR}}
solutions will accordingly yield dissimilar descriptions of the  
problematic hot quantum field theories IR sectors
[8,13,15].\smallskip

 This point of view is
supported also by some recent calculations [14] performed in the  
context of the {\it{collinear singularity}}
problem of hot QCD [15], with most encouraging results. For the  
moving fermion damping rate problem we have taken
as an illustration throughout this letter, we see that the same IR  
diverging result as found in the {\it{RP}}
approach, is eventually derived in the {\it{PR}} scheme, at the  
same fermionic mass shell limit. In
contradistinction with the {\it{RP}} calculation, this result is  
exclusively due to the longitudinal degrees of
freedom, whereas transverse ones do not contribute. Now, along the  
{\it{RP}}, there is effectively no buit-in
argument to avoid the dangerous low $k's$ IR domain
of integration, whereas {\it{Perturbative Resummations}} can  
certainly invoke the finite convergence
radius of their geometrical series.

 Now, this procedure also begs the
question of completeness, because there is no guarantee that the  
full leading order fermionic damping rate is
obtained. However, this flaw is not decisive either, so as to  
discriminate between {\it{RP}} and
{\it{PR}} resummation
procedures, firstly, because the {\it{RP}} is itself an incomplete  
scheme in general [13], and in second place,
because there is no guarantee either, that the $RP$ be duly used,  
when used at arbitrarily soft momenta! Though
rather natural an idea indeed, it is only now that the latter comes  
in view in the literature [16], in obvious
connection with the breakthrough of order $g^2T$ effective theories  
[17].\smallskip
Unfortunately, we are not yet in a situation where some  
{\it{experimental numbers}} could help us knowing how far
we stand from a satisfying
description of the ``thermal quantum world", though recent  
estimates would seem to revive some hope in this
direction [18].

\vfill\eject

[1] G.H. Hardy, ``Divergent Series"  (Oxford Clarendon Press,  
1949).  \smallskip\smallskip\smallskip

[2] G. Grunberg, preprint hep-ph/0009272  \smallskip\smallskip\smallskip

[3] T.
Rado\.zycki, {\it{Eur. Phys. J. C}} {\bf{6}}, (1999) 549  
.\smallskip\smallskip\smallskip

[4] B. Candelpergher and T. Grandou, {\it{Ann. Phys. (N.Y.)}}  
{\bf{283}}, (2000) 232.
\smallskip\smallskip\smallskip

[5] M. Le Bellac, ``Thermal Field Theory" (Cambridge University
Press, 1996). \smallskip\smallskip\smallskip

[6] J. Bros and D. Buchholz, {\it{Ann. Inst. Henri Poincar\'e}}, Vol.
{\bf{64}}, (1996) 495.\smallskip\smallskip\smallskip

[7] R. Kobes, G. Kunstatter and A. Rebhan, {\it{Nucl. Phys.
B}}{\bf{355}}, (1991) 1. \smallskip\smallskip\smallskip

[8] R.D. Pisarski, {\it{Phys. Rev.
Lett.}}{\bf{63
}}, (1989) 1129. \smallskip\smallskip\smallskip

[9] J.P. Blaizot, E. Iancu
and R. Parwani, {\it{Phys. Rev. D}} {\bf{52}}, (1995) 2543.  
\smallskip\smallskip\smallskip

[10] D. Karabali and V.P. Nair, {\it{Nucl. Phys.
B}}{\bf{464}}, (1996) 135.\smallskip J. Reinbach and H. Schulz,  
{\it{Phys. Lett. B}} {\bf{467}}, (1999)  
247.\smallskip\smallskip\smallskip

[11] P. Aurenche, F.
Gelis and H. Zaraket, preprint LAPTH-756/99, BNL-NT-99/5  
\smallskip\smallskip\smallskip

[12] T. Grandou,
{\it{Phys. Lett. B}} {\bf{367}}, (1996) 229;\smallskip T. Grandou  
and P. Reynaud,
{\it{Nucl. Phys. B}}{\bf{486}}, (1997) 164.\smallskip\smallskip\smallskip

[13] P. Aurenche, F.
Gelis, R. Kobes and H. Zaraket,\smallskip
{\it{Phys. Rev. D}} {\bf{58}}, (1998) 085003.\smallskip P. Aurenche, F.
Gelis, R. Kobes and E. Petitgirard, {\it{Z. Phys. C}} {\bf{75}},  
(1997) 315.
\smallskip\smallskip\smallskip

[14] T. Grandou, preprint INLN 2000/27.\smallskip\smallskip\smallskip

[15] R. Baier, S. Peign\'e and D. Schiff,
{\it{Z. Phys. C}} {\bf{62}}, (1994) 337.\smallskip\smallskip\smallskip

[16] S. Leupold and M. Thoma, {\it{Phys. Lett. B}} {\bf{465}},  
(1999) 249.
\smallskip\smallskip\smallskip

[17] E. Iancu, {\it Phys. Lett. B} {\bf 435} (1998) 152; D.  
B\"odeker, {\it Phys. Lett. B}
{\bf 4}26 (1998) 351;\smallskip P. Arnold, D. Son and L.G. Yaffe,  
{\it{Phys. Rev. D}} {\bf{55}} (1997)
6264.\smallskip\smallskip\smallskip

[18] R. Baier, A.H. Mueller, D. Schiff and D.T. Son, preprint  
hep-ph/0009237.

\end